\documentclass[twocolumn,prb,showpacs,amsmath,amstex,amssymb,mathfonts,superscriptaddress]{revtex4-1}

\usepackage{amsthm,color,amsfonts,graphicx,verbatim}
\usepackage[dvipsnames]{xcolor}
\usepackage{amsmath}
\usepackage{amssymb}
\usepackage{amsthm}
\usepackage{amsfonts}
\usepackage{listings}
\usepackage{enumerate}
\usepackage{latexsym}
\usepackage{natbib}
\usepackage{epstopdf}
\usepackage{epsfig}
\usepackage{import}
\usepackage{psfrag}
\usepackage{bm}
\usepackage{graphicx}
\usepackage{dsfont}
\usepackage[normalem]{ulem}
\usepackage{braket}

\newcommand{\rl}{\mathrm{l}}
\newcommand{\rr}{\mathrm{r}}
\newcommand{\rE}{\mathrm{E}}

\newcommand{\be}{\begin{equation}}
\newcommand{\ee}{\end{equation}}
\newcommand{\beq}{\begin{equation}}
\newcommand{\eeq}{\end{equation}}
\newcommand{\bea}{\begin{eqnarray}}
\newcommand{\eea}{\end{eqnarray}}

\newcommand{\e}{\mathrm{e}}

\renewcommand{\phi}{\varphi}
\renewcommand{\epsilon}{\varepsilon}
\newcommand{\str}{ |}
\newcommand{\norm}{ ||}

\def\nn{\nonumber\\ }

\newcommand{\loct}{\zeta}
\newcommand{\loca}{\bar{\zeta}}
\newcommand{\xia}{\bar{\xi}}

\newcommand{\rddensity}{\rho}

\usepackage{hyperref}
\hypersetup{
    bookmarks=true,         
    unicode=false,          
    pdftoolbar=true,        
    pdfmenubar=true,        
    pdffitwindow=false,     
    pdfstartview={FitH},    
    pdftitle={Many-body delocalization as a quantum avalanche},    
    pdfauthor={},     
    pdfsubject={},   
    pdfcreator={},   
    pdfproducer={}, 
    pdfkeywords={keyword1} {key2} {key3}, 
    pdfnewwindow=true,      
    colorlinks=true,       
    linkcolor=black,          
    citecolor=blue,        
    filecolor=magenta,      
    urlcolor=blue           
}

\begin{document}
\title{Many-body delocalization as a quantum avalanche}

\author{Thimoth\'ee Thiery}
\email{{thimothee.thiery@kuleuven.be}}
\affiliation{Instituut voor Theoretische Fysica, KU Leuven, Belgium}

\author{Fran\c{c}ois Huveneers}
\email{huveneers@ceremade.dauphine.fr}
\affiliation{Universit\'e Paris-Dauphine, PSL Research University, CNRS, CEREMADE, 75016 Paris, France}

\author{Markus M\"uller}
\email{{markus.mueller@psi.ch}}
\affiliation{Paul Scherrer Institute, PSI Villigen, Switzerland}

\author{Wojciech De Roeck}
\email{{wojciech.deroeck@kuleuven.be}}
\affiliation{Instituut voor Theoretische Fysica, KU Leuven, Belgium}

\date{\today}
\begin{abstract}
We propose a multi-scale diagonalization scheme to study disordered one-dimensional chains, 
in particular the  transition  between many-body localization (MBL) and the ergodic phase, expected to be governed by resonant spots.
Our scheme focuses on the dichotomy MBL versus {ETH} (eigenstate thermalization hypothesis). 
We show that a few natural assumptions imply that the system is localized with probability one at criticality.
On the ergodic side, delocalization is induced by a quantum avalanche seeded by large ergodic spots, whose size diverges at the transition. 
On the MBL side, the typical localization length tends to a finite universal value at the transition, 
but there is a divergent length scale related to the response to an inclusion of large ergodic spots.   
A mean field approximation analytically illustrates these results and predicts as a power-law distribution for thermal inclusions at criticality.

\end{abstract}

\maketitle

\textbf{Introduction} ---
The phenomenology and theory of MBL, i.e.\@ the absence of thermalization in interacting quantum systems\cite{anderson_absence_1958,fleishman_interactions_1980,basko2006metal,gornyi2005interacting,znidaric_many-body_2008,oganesyan_localization_2007,pal_many-body_2010,ros2015integrals,kjall2014many,luitz_many-body_2015,nandkishore_many-body_2015,altman_universal_2015,abanin_recent_2017,luitz_ergodic_2017,agarwal_rare-region_2017,imbrie_review:_2016}, challenges our understanding of statistical mechanics.  
In $d=1$, the main outstanding issue is the nature of the transition\cite{grover2014certain,serbyn2015criterion,potter2015universal,vosk2015theory,khemani2017critical,PhysRevLett.119.075702,PhysRevB.96.104205,kulshreshtha2017behaviour,parameswaran_eigenstate_2017,PhysRevB.93.224201,PhysRevLett.119.110604,PhysRevLett.115.187201,PhysRevB.94.045111} that separates the MBL from the ergodic (thermalizing) phase.  
To describe it, several phenomenological renormalization schemes have been introduced \cite{potter2015universal,vosk2015theory,PhysRevB.93.224201,PhysRevLett.119.110604}, with partially conflicting predictions.

In the present Letter, we develop a theory that is rooted in two microscopic principles. 
The first principle, governing non-resonant couplings, is spectral perturbation theory. 
The second principle is the use of random matrix theory for resonant couplings\cite{deutsch1991quantum,srednicki1994chaos,rigol_thermalization_2008,d2015quantum,chandran2016many,ponte2017thermal,de2017stability}, 
which strikingly predicts an `avalanche' instability: 
An infinite localized system can be thermalized by a finite ergodic seed if the typical localization length $\zeta$ exceeds a critical $\zeta_c$\cite{de2017stability}. 
We implement these principles in the form of a multi-step diagonalization procedure\cite{wegner1994flow,imbrie2016many,PhysRevLett.116.010404,PhysRevLett.119.075701,monthus2016flow,rademaker2017many}, described compactly below and in more details in the companion paper\cite{thimotheelong}. 

Analyzing first the general consequences of this scheme, we find that the critical point must be localized with probability one. 
This conclusion, that rests on a few basic facts and does not involve any detailed analysis,
contrasts with predominantly numerical RG studies \cite{potter2015universal,PhysRevLett.119.110604} 
where the half-chain entanglement entropy at the critical point was reported to follow a (sub-thermal) volume law. 
It implies that the entanglement entropy associated with typical cuts is discontinuous at the transition, as in \cite{khemani2017critical}, 
and that the typical localization $\loct$ does not diverge.  

\begin{figure}
\centerline{\includegraphics[width=8cm]{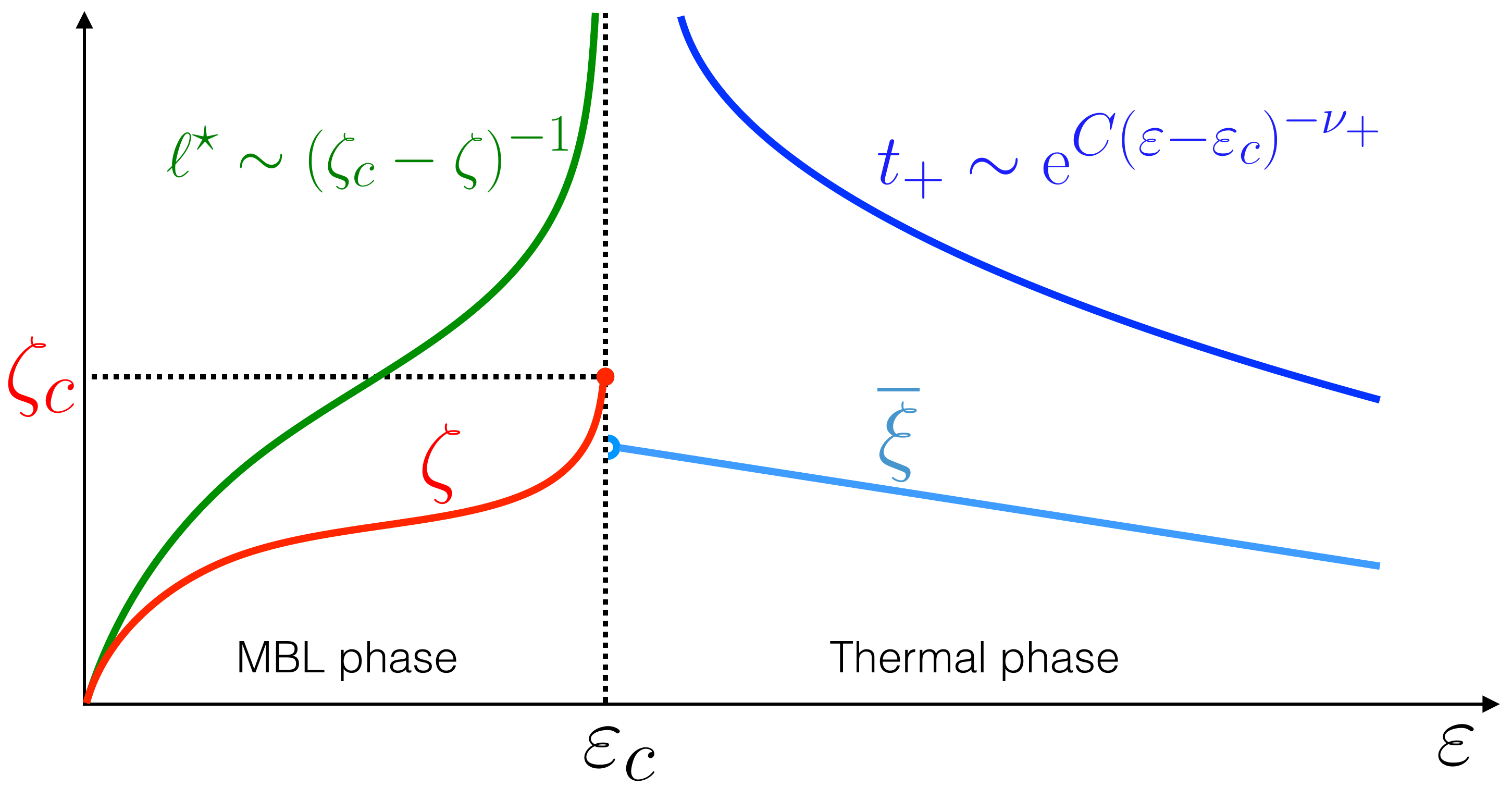}} 
\caption{
Phase diagram where $1/\epsilon$ quantifies the disorder strength. 
$\loct$ is the typical localization length. 
It is bounded by the critical localization length $\zeta_c$, which equals the inverse of the entropy density. 
$\ell^\star$ quantifies the susceptibitlity to the insertion of large ergodic spots.
$\bar\xi$ is the average eigenstate correlation length. 
$t_+$ is a typical local thermalization time scale in the thermal phase.
}
\label{Fig:Phasediagram}
\end{figure}

This last point is a direct consequence of the explicit upper bound $\loct \leq \zeta_{c}$\cite{de2017stability}.
On the other hand, we do identify a length scale $\ell^\star$ that does diverge as $(\loct-\loct_c)^{-1}$ as one approaches the transition from the MBL side, 
see Fig.~\ref{Fig:Phasediagram}.  
This is caused by the divergent susceptibility of the sample to the insertion of large ergodic spots. 
In our scheme such spots trigger delocalization by an avalanche instability, 
a central aspect that distinguishes our work from previous approaches\cite{vosk2015theory,potter2015universal,PhysRevB.93.224201,PhysRevLett.119.110604}.
The validity of the avalanche scenario and the associated bound on $\loct$ were recently verified through high precision numerics\cite{luitz2017}, 
and we show here that it leads to a consistent and physical picture of the MBL transition. 
On the thermal side instead, we find no divergent correlation length, but only a diverging cross-over length $L_+$, beyond which typical samples appear thermal. 
$L_+$ is associated with a typical time scale of local thermalization, $t_+$, that diverges quasi-exponentially at the transition. 

We illustrate these aspects with a mean-field approximation of our scheme. 
While it introduces some over-simplifications, as discussed below, it offers a very concrete implementation of the main ideas developed in this Letter, 
and yields several conclusions that have been confirmed by a full numerical analysis of the scheme\cite{thimotheelong}.
These includes the fact that the upper bound $\zeta_c$ is indeed saturated at the transition, as well as a power-law distribution for thermal inclusions at criticality.

\textbf{Multi-step diagonalization scheme ---} 
We consider a {chain of spins $S_i$, with $d_s$ states per site $i$}, with a generic local Hamiltonian, which we write in the form
\begin{equation}\label{eq: original Hamiltonian}
	H= \sum_{I} D_I + \sum_{I} V_{I}.
\end{equation}
Here, $I$ denotes a stretch of consecutive sites that the operators $D_I,V_I$ act on. 
The operators $D_I$ act diagonally in a preferred basis that for concreteness, we take to be the $S^z_i$-basis. 
The $V_I$ are not diagonal and we refer to them as `couplings'. We express lengths in units of the lattice spacing $a$. 
A special role is played by the entropy density $s = \log (d_s) / a$. 
We now diagonalize the system by iteratively eliminating couplings, see~Fig.~\ref{Fig:FirstBlob}.

\paragraph{Perturbative couplings.}
The distinction between perturbative and resonant couplings is at the heart of our procedure. 
We declare a coupling $V_I$ `perturbative' if  {typical}  eigenstates of ${D}_I+V_I$
are small perturbations of the eigenstates of ${{D}}_I$ and can hence be obtained by perturbation theory.  
Following \cite{imbrie2016many}, 
we prefer to think in terms of a unitary transformation $U_I$ that eliminates the coupling $V_I$ to lowest order by acting on $H$ as $H \to U_I H U_I^\dagger$. 
This is achieved by choosing {$U_I=\e^{A_I}$ with
$
 \langle \eta' \str A_I \str  \eta \rangle :=  \frac{\langle \eta' \str  V_I \str  \eta \rangle  }{E_I(\eta')-E_I(\eta)} .
$
Here, $\eta,E_I(\eta)$ are eigenstates and eigenvalues of $ D_I$. This procedure is meaningful if 
\begin{equation}\label{eq:resonancecondition}
	\mathcal G \equiv  \max_{\eta'}\left\str {\langle \eta' \str A_I \str  \eta \rangle  } \right\str <1 \quad\text{for typical $\eta$,} 
\end{equation}
see supplemental material (SM).
If $V_I$ is not perturbative, i.e.\@ if $\mathcal G \ge 1$, then we call the coupling $V_I$ resonant and we do \emph{not} eliminate it. 
By eliminating a perturbative coupling, we generate new, but usually smaller couplings. Indeed, whenever $I\cap J \neq \emptyset$, we will create a new coupling $V_{I'}=U_IV_JU_I^{-1}-V_J\approx  [A_I,V_J] $ (at first order) with $I'\equiv I\cup J$.  

\paragraph{Resonant couplings.}
We first eliminate all perturbative couplings that do not touch resonant regions, see Fig.\@~\ref{Fig:FirstBlob}.
\begin{figure}
	\centerline{
	\includegraphics[width=9cm, height=3.5cm]{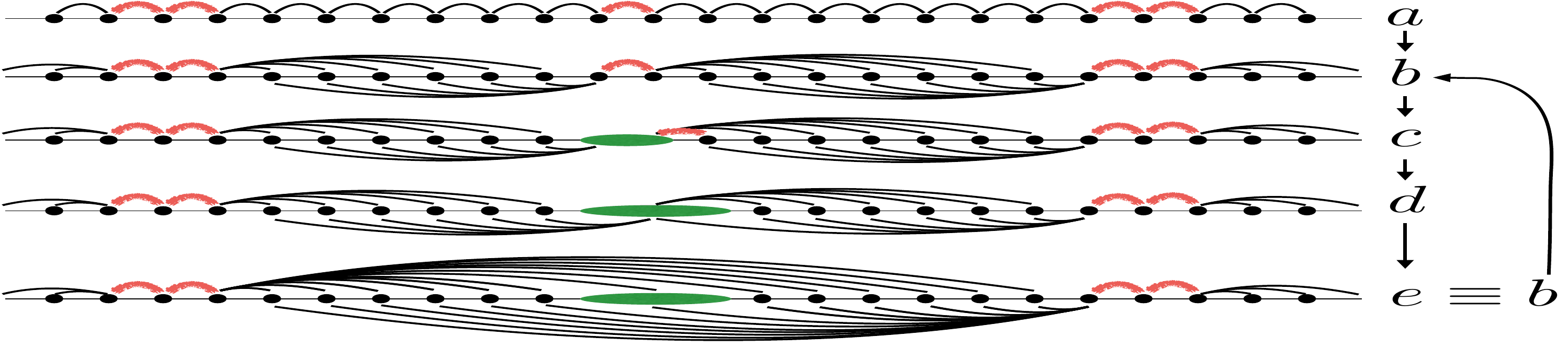}}
	\caption{Evolution of the spin chain during the diagonalization procedure. 
	Black dots are spins and the green ellipse is a group of non-perturbatively diagonalized spins. 
	Arcs symbolize couplings that act on all spins they embrace. Red ones are resonant, black ones are perturbative.   
	$a)$: Initial nearest-neighbor couplings.  
	$b)$: All perturbative couplings that do not touch resonant couplings have been eliminated, which generates weak couplings to the adjacent resonant spots. 
	$c)$: Some resonant sites have been fused into a green spot, requiring a re-evalution of the existing couplings. 
	Some arcs to sites close to the green spot have thereby become resonant (red).   
	$d)$: The new resonant coupling is fused into the green spot. No new red couplings emerged. 	
	They are eliminated in step $e)$. 
	A next resonant spot can be fused now, as in $b)$, until no resonances are left.}
	\label{Fig:FirstBlob}
\end{figure} 
We then assume that the remaining resonant couplings induce full ergodicity locally, i.e.\@ within their range. 
We thus diagonalize them by a unitary that we consider as an effective random matrix.  
Such a strong dichotomy has been theoretically predicted\cite{de2017stability} and these predictions are in remarkable agreement with numerics \cite{luitz2017}.
The random matrix ansatz remains consistent throughout the scheme if perturbation theory is used as much as possible to `isolate' resonances from their environment\cite{de2017stability,imbrie2016many}. 
For this reason we diagonalize resonances in order of increasing size and leave untouched perturbative couplings that link to ergodic spots~\cite{thimotheelong,ordernote}. 
The diagonalization of a resonant region alters the perturbative couplings attached to it and can turn them resonant (and thereby potentially start an avalanche). 
If they remain perturbative they can be eliminated in the next step as the scheme is iterated, see Fig.~\ref{Fig:FirstBlob}. 
After a number of iterations which scales logarithmically with system size,  all couplings will be eliminated.

\paragraph{End of the procedure.}
The final result of our procedure is encoded in the diagonalizing unitary  $U$. It is obtained as a product of unitaries $U_{I_n}$, each acting on a single stretch $I_n$, as described above. 
A region  $Y$ is ergodic if $Y=\cup_\gamma I_\gamma$ for a collection of intersecting stretches $\{I_\gamma\}$ and such that all $U_{I_\gamma}$ are non-perturbative. 
The full system is thermal if the whole sample is ergodic, otherwise it is by definition MBL.
From $U$, one obtains local integrals of motion\cite{imbrie2016many,PhysRevLett.111.127201,PhysRevB.90.174202,huse2014phenomenology} (LIOMs) by inverse conjugation:
\begin{equation}\label{eq: definition tau}
\tau_i =      U^\dagger S^z_i U.
\end{equation}
We decompose $\tau_i=\sum_{I}  \tau_{i,I}$ in spatial components, where $\tau_{i,I}$ acts on the stretch $I$. 
The decay of $\tau_{i,I}$ with increasing $| I |$ defines the typical localization length:  
\begin{equation} \label{def: loc lengths}
	\loct^{-1}= -\lim_{|I| \to \infty} \frac{\langle {\log \norm  \tau_{i,I} \norm } \rangle }{|I| },
\end{equation}
where  $\norm \cdot \norm$ is the operator norm\cite{notenorm}, whose crucial role will be explained below, 
and where $\langle \cdot \rangle$ denotes the disorder average, assuming that some randomness enters in the original Hamiltonian in Eq.~\eqref{eq: original Hamiltonian}.
The locality properties of the unitary $U$ also reflect in the spatial decay of couplings $V_I$ ({created and eliminated in the course of our scheme}), 
as they arise from a transformation inverse to \eqref{eq: definition tau}. 
$\loct$ can thus also be defined by replacing $\tau_{i,I}$ in \eqref{def: loc lengths} by $V_I$.

\setcounter{paragraph}{0}

{\textbf{General analysis of the scheme ---}  
For any disorder realization, at a finite length $L$, let $\rho_{\rm th}(L)$ be the density of sites in an ergodic region, as determined by the scheme.  
Let us now assume that our scheme has the two following properties. 
(i) Denoting the inverse disorder strength by $\varepsilon$ (see below), 
we assume that $\langle \rddensity(L) \rangle_\varepsilon$ is continuous and non-decreasing as $\varepsilon$ increases.
(ii) For any given disorder realization, all sites declared thermal remain so if more sites are added to the chain at one or both ends. 
{One should probably not expect (ii) to hold strictly (e.g.\ coupling a site to a very disordered region might indeed increase the effective disorder on that site, see also \cite{thimotheelong}). Yet, it seems to us that these properties are intuitive and reasonable properties for phenomenological models. }}

{We first deduce that $\langle \rddensity (L) \rangle_\varepsilon$ reaches a limit $\rddensity^\star(\varepsilon)$ as $L \to \infty$, for any $\varepsilon$. 
Indeed, (ii) directly implies the superadditivity property
$
	\rddensity (L+L') \ge \frac{L}{L+L'} \rddensity (L) + \frac{L'}{L + L'} \rddensity(L')
$
and then the limit exists by Fekete's superadditivity lemma. 
Second, we show a concentration property: 
For any disorder strength $\varepsilon$ and for all but a vanishing fraction of the samples, the thermal density approaches $\rddensity^\star(\varepsilon)$, 
i.e.\@ $P(|\rddensity(L) - \rddensity^\star (\varepsilon) | > \delta)\to 0$ for any $\delta > 0$ as $L \to \infty$. 
Since this property is valid in particular at the critical point $\epsilon = \epsilon_c$, 
it implies that it must be either localized or thermal with probability 1. 
To see the concentration of $\rddensity$, 
let us fix $\delta > 0$ and let $L_0$ be large enough so that $|\langle \rddensity (L_0) \rangle_\varepsilon-\rddensity^\star (\varepsilon) | < \delta/2$. 
Let us then consider a `product' system made of blocks of size $L_0$ that are decoupled from each other. 
(ii) implies that 
$
	P(\rddensity(L) > a ) \ge P_{\rm prod}(\rddensity(L) > a). 
$
The concentration property that we seek holds definitely true for the product system.
We thus conclude that $P(\rddensity (L) - \rddensity^\star(\varepsilon) < - \delta)$ goes to $0$ as $L \to \infty$ and, 
since the average value of $\rddensity$ converges to $\rddensity^\star (\varepsilon)$, 
we also conclude that $P (\rddensity - \rddensity^\star (\varepsilon) > \delta)$ must vanish as $L \to \infty$. }

{It remains to decide whether the critical point is localized or thermal. 
Since $\langle \rddensity (L) \rangle_\varepsilon$ is non-decreasing in both $L,\epsilon$ and continuous in $\epsilon$ by (i) and (ii), 
we directly conclude that  $ \rddensity^\star (\varepsilon)$ is non-decreasing and left-continuous. 
Hence, either $\rddensity^\star (\varepsilon)$ is actually continuous at the transition, and the critical point is thermal, 
or it has a jump at the transition and the critical point is localized. 
It is clear that the localization length $\zeta$ should diverge as $\rddensity^\star$ approaches 1 
(cf.\@ the `rule of halted decay' below), 
hence the bound $\zeta \le \zeta_c$\cite{de2017stability,luitz2017} implies that $\rho_c < 1$ and thus that the critical point is localized. }

\textbf{Understanding the transition ---}
{We now develop a general picture for the transition. 
By making a simple mean-field assumption, we also get explicit analytical results substantiting this picture.  
Detailed numerics in \cite{thimotheelong} yields further support for our theory.
}

\paragraph{Resonances and scales}
The simplest resonances are  associated to couplings $V_I$ on single bonds $I=\{i,i+1\}$. 
Let $\epsilon$ be the probability that such a $V_I$ is resonant and let us use this as a measure for the inverse disorder strength. 
We call a \emph{bare spot} of order $k$ a set of $k$ adjacent resonant bonds. 
The density of such bare spots is $\epsilon^k/a$, and their distribution is the only randomness taken into account: 
the localized material between these spots is homogeneous with bare localization length $\zeta_1$. 
In other words, we consider a bimodal distribution of nearest-neighbour couplings.
We  {parametrize}\cite{thimotheelong} $\zeta_1(\epsilon) = - 1/\log(\epsilon/K)$ with $K$ a non-universal constant.  
In our scheme, we treat the smallest resonant spots first and thus it is natural to think of the order $k$ as  an effective scale.
We introduce the {\em running} localization length $\loct_k$ as above, but using a unitary $U$ that only eliminates the spots of order $k' < k$ 
(alternatively, replacing $|I|\to\infty$ by $|I|\sim a \epsilon^{-k}$). 
$\loct_k$ is thus the effective localization length seen by spots of order $k$. 
It increases with $k$, since increasingly more effects of resonant spots are included.
Indeed, a calculation yields the important `rule of halted decay': 
If a fraction $\rho_I$ of a stretch $I$ is thermal, then one finds  $\norm V_I \norm \sim \e^{-(1-\rho_I)|I|/\loct_1}$
for couplings $V_I$ that are relevant to our scheme, see SM.

A spot of order $k$ melts (or thermalizes) a region of length $\ell_{k}$ on each side of the bare spot. 
The fraction of space occupied by such thermal regions is $\rho_k\equiv\epsilon^k (k+2\ell_k/a)$.  
Hence $\rho_k$ is the density of additional thermalized regions that has to be accounted for when passing from $\loct_k$ to $\loct_{k+1}$. 
How precisely $\loct$ is assumed to increase does not affect the resulting key features. 
Using the rule of halted decay, a simple possibility is the mean field approximation (see SM)
\begin{equation} \label{eq: populistic}
\loct_{k+1}^{-1}= (1-\rho_k)\loct_{k}^{-1}.
\end{equation}

\paragraph{Large resonant spots}
We now derive an expression for the length of the melted region $\ell_k$, in agreement with the theory developed in\cite{de2017stability,luitz2017}.
The couplings linking a bare spot of order $k \gg 1$ to its close surroundings are typically resonant in the early iterations of the scheme, 
see Fig.~\ref{Fig:spot} in the SM. 
After thermalizing $\ell$ spins on each side, the couplings $V_{\mathrm{E-l}}, V_{\mathrm{E-r}}$ from the spot to the spins $\mathrm{l},\mathrm{r}$ just outside it, 
originate microscopically from the (by now rotated) couplings $\widetilde V_{\mathrm{E-l}}, \widetilde V_{\mathrm{E-r}}$ 
between the spins $\mathrm{l},\mathrm{r}$ and the peripheral spins of the bare spot.  
Those scaled as
$
	\norm \widetilde V_{\mathrm{E-l}} \norm,  \norm \widetilde V_{\mathrm{E-r}} \norm \sim g_0 e^{-\ell/\loct_k}
$. 
Since we have diagonalized the spot by a random unitary, any structure  distinguishing the coupling operators from random matrices (i.e., ETH behavior) has been erased, 
but the norm of the operators is preserved. 
Hence we know that 
$
\norm  V_{\mathrm{E-l,r}} \norm= \norm \widetilde V_{\mathrm{E-l,r}} \norm .
$ 
As this coupling is now indeed a random matrix acting on a space of dimension $d_{\ell}\equiv\e^{s(ak+2\ell)}$ (since the spot has grown from two sides), 
its matrix elements have size $g_0\e^{-\ell/\loct_k} d^{-1/2}_{\ell}$. 
Hence the condition to be perturbative is 
$
 \e^{s(ak+2\ell)/2}\e^{-\ell/\loct_k} \leq 1 .
$ 
Thus, spins are thermalized up to distance $\ell=\ell_k$ with
\begin{equation} \label{eq: def lc}
	\ell_k = k \tfrac{s a}{2} \left(\tfrac{1}{\loct_k}-s\right)^{-1}.
\end{equation}
Since $\loct_k\to \loct$ as $k\to \infty$, but $\ell_k< \infty$ in the MBL phase, we derive a bound on the typical localization length
\begin{equation} \label{eq: bound on zetatyp}
	\loct \leq  \zeta_c\equiv  s^{-1}.
\end{equation}
If this bound were violated, a finite spot would trigger an avalanche and delocalize an arbitrarily large system.   {As we have seen above, a system appears less localized at larger scales. The picture that emerges is thus that  $\lim_{k\to\infty}\zeta_k=s^{-1}$ at the transition, which is hence driven by infinite spots}. 

\paragraph{Discussion}
From the relation \eqref{eq: def lc} and assuming the recursion equation \eqref{eq: populistic}, we can render the flow of $\loct_k$ near the transition explicit, see SM. 
We find a transition at $\epsilon=\epsilon_c \in ]0,1[$ defining three regimes:\\
(i)\emph{Localized regime} $\epsilon<\epsilon_c$. 
At large scale, $\zeta_k\to \zeta<\zeta_c$ and, from Eq.~\eqref{eq: def lc},
\begin{equation}\label{eq: def l star}
	\ell_k / k \; \to \;  \ell^\star \,\sim \, (\zeta_c - \zeta)^{-1},   
\end{equation}
as $k \to \infty$,
where $\ell_k/k$ represents the susceptibility of the material to the insertion of a bare thermal spot of size $k$. 
We expect {$\zeta_c-\zeta\sim (\epsilon_c-\epsilon)^{\nu_-}$}, leading to $\ell^\star \sim (\epsilon_c- \epsilon)^{-\nu_-}$. 
Numerics yields $\nu_-\in [1/2,1]$ to be nonuniversal: It depends on the parameter $K$. 
Moreover, $\ell^\star$ is the scale at which the power law distribution of thermal spot sizes is cut off exponentially, see SM.\\
(ii)\emph{Critical regime} $\epsilon=\epsilon_c$. 
At large scale  $\zeta_k\to \zeta_c$  and the bound \eqref{eq: bound on zetatyp} is saturated.
From Eq.~\eqref{eq: def lc}, the susceptibility $\ell_k/k$ diverges thus as $k\to\infty$ ($\ell_k \sim \epsilon_c^{-k/2}$).
Yet, the system is localized and the thermal density $\rddensity$ is strictly smaller than $1$. 
The probability of having a thermal region of size $\ell$ centered on a given site scales as $p(\ell)\sim \ell^{-\tau}$, with $\tau=3$. 
While the {\it typical} half-chain entanglement entropy $S$ is {hence} bounded, 
its {\it average} (over samples) diverges as  $\overline{S} \sim \int^{L} \ell^2 p(\ell) d\ell \sim \log(L)$ with system size $L$.\\
(iii)\emph{Thermal regime} $\epsilon >\epsilon_c$. 
At small scales, the system appears localized ($\zeta_k <\zeta_c$) but at a finite $k_*$, $\zeta_{k_*}\geq \zeta_c$ implying $\ell_{k_*}=\infty$ and triggering an avalanche. 
The critical core size $k_*$ diverges logarithmically $k_* \sim \nu_+ \frac{\log(\epsilon-\epsilon_c)}{\log(\epsilon_c)}$ with $\nu_+ >0$.

\setcounter{paragraph}{0}
\textbf{Finite-size scaling ---} 
Let us evaluate the probability $p(\epsilon,L)$ of a chain of length $L$ to be thermal. 
For large $L$, to exponential accuracy, 
we find $p(\epsilon,L)= \epsilon^{L/L_-}$ with $L_- \sim |\epsilon- \epsilon_c|^{-\nu_-}$ in the MBL phase (by requiring a thermal spot to cover the whole system).
On the thermal side, the system is ergodic unless it contains no explosive spots. 
This yields $p(\epsilon,L)= 1- (1-\epsilon^{k_*})^{L} \approx 1-\exp(-L/L_+)$ with $L_+= \epsilon^{-k_*}\sim |\epsilon- \epsilon_c|^{-\nu_{+}}$, 
with a non-universal $\nu_+\in[1,2]$.
In the critical fan, one finds  $p(\epsilon,L)\sim L^{-\beta}$ with ${\beta}=\tau-2=1$, 
as follows from the estimate $p(\epsilon,L) \sim L\int_{\ell > L } p(\ell) d\ell$. 
{In particular, close above the transition point, $p(\varepsilon,L)$ has a non-monotonic behavior as a function of $L$: 
It first decreases polynomially and then grows to 1 exponentially fast.}

\setcounter{paragraph}{0}
\textbf{Correlation lengths ---}
The MBL transition does not manifest itself in thermodynamic correlation functions, but only in dynamic properties such as eigenstate correlation functions like
\begin{equation} \label{def: correlation}
	\mathrm{Cor}(O_0,O_\ell) \equiv \overline{
\str\langle \Psi, O_0   O_\ell  \Psi \rangle-  \langle \Psi, O_0  \Psi \rangle  \langle \Psi,  O_\ell  \Psi \rangle \str},
\end{equation}
for local operators $O_i$ acting around site $i$, where the average is over both $\Psi$ and disorder.  Numerical studies \cite{ZhangKhemaniHuse2016,rademaker2017many} had suggested that 
\begin{equation}\label{eq:defcorrelationlength}
	\xia^{-1} =  \lim_{\ell\to \infty}\frac{1}{\ell}\log { \mathrm{Cor}(O_0,O_\ell)}
\end{equation}
diverges at the transition, while being finite in both bulk phases. 
Indeed, in the ergodic phase, by the ETH, the quantity \eqref{def: correlation} equals the thermodynamic correlator, which decays exponentially.  
Let us now determine $\xia$.

\paragraph{MBL phase}
A detailed analysis (SM) yields the following picture: 
If $O_0,O_\ell$ are located on either side of, but just outside, a thermal region (bare spot + full melted region), then $ \mathrm{Cor}(O_0,O_\ell) \sim 1$.  
This is easiest understood by realizing that the couplings connecting the spot to sites $0$ or $\ell$ are barely non-resonant,  
because those sites are just outside the melted regions. 
This self-organized criticality affects the diagonalizing unitaries, yielding $ \mathrm{Cor}(O_0,O_\ell) \sim 1$. 
Taking into account the probability of thermal regions, we conclude that $\xia\sim \ell^\star$. 

Note the following counterintuitive implication: If we increase the couplings that link the thermal region to spins $0,\ell$, such that these spins melt as well, the $\mathrm{Cor}(O_0,O_\ell)$ is \emph{lowered}, as it is now computed in a thermal system. This is reminiscent of the principle of \emph{monogamy of entanglement}.

\paragraph{Thermal phase}
It is often suggested that $\xia$ as defined in \eqref{eq:defcorrelationlength} also diverges from the thermal side. However, within our scheme an infinite chain on the thermal side is thermal with probability $1$, so that ETH applies and the correlator \eqref{def: correlation} thus decays exponentially in $\ell$.
Yet, if $\xia$ is defined\cite{ZhangKhemaniHuse2016} via Eq.~\eqref{eq:defcorrelationlength} with local operators acting on the two ends of the chain,
then, with probability $\e^{-L/L_+}$, the correlator probes a localized chain and it can hence fail to decay. 
By exhibiting (SM) a mechanism for long-range correlations, we put  the lower bound $(\log L_+)^{-1} \sqrt{L_+}$ on the associated divergent correlation length. 

\textbf{Critical slowing down ---}
While there is no natural diverging correlation length on the thermal side, there is definitely a diverging time scale $t_+$: the inverse of local thermalization rates for typical spins. 
Indeed, if a spin is eventually thermalized by a thermal region emanating from a bare spot at distance $\ell$, the Fermi Golden Rule roughly yields a flipping rate $e^{-2\ell/\loct} \Gamma_0$ with  a microscopic rate $\Gamma_0$.  In the thermal phase most spins will be thermalized by a bare spot of order $k_*=k_*(\epsilon)$, located at a typical distance $\epsilon^{-k_*}\sim L_+ \sim (\epsilon-\epsilon_c)^{-\nu_+}$. 
Therefore, $ t_+ \sim e^{C (\epsilon-\epsilon_c)^{-\nu_+}} $, diverges {(quasi)}-exponentially. 
While this result concerns typical sites in the chain, transport over long distances $L \gg L_+$ is dominated by rare localized Griffiths regions~\cite{lev2015absence,luitz2016extended,gopalakrishnan2016griffiths,luitz2017ergodic,vznidarivc2016diffusive,nahum2017dynamics},
realized here as regions containing no explosive spots with $k\geq k_*$. This leads (SM) to the subdiffusive scaling 
$R(L) \sim L^{C(\epsilon-\epsilon_c)^{-\nu_+}}$ of resistance $R=R(L)$ close to the transition.

\textbf{Beyond mean-field ---} 
We emphasize that the mechanism for the transition studied above through a mean field approximation in Eq.~\eqref{eq: populistic},
together with its consequences, such as the divergent susceptibility $\ell^\star$, 
saturation of the bound \eqref{eq: bound on zetatyp}, the divergence of the correlation length $\overline{\xi}$, etc$\dots$
were obtained just by assuming the consistency of a scale-by-scale analysis, without relying on the specific flow equation \eqref{eq: populistic}. 
These conclusions are hence confirmed by the numerical analysis\cite{thimotheelong} of our full non-simplified scheme. 
The most important new feature that emerges there is the fact that ergodic spots of low order $k$ arranged close to each other in a fractal pattern,
have the same delocalizing power as a spot of much higher order. 
This means that the concept of `bare ergodic spot' should be refined. 
As a consequence, the tail of the subcritical probability $p(\ell)$ of thermal spots is a stretched exponential rather than an exponential. 
Hence $\overline{\xi}$ is \emph{strictu sensu} always infinite in the localized phase. 
However, upon replacing $\frac{1}{\ell}$ by $\frac{1}{\ell^b}$ in the definition \eqref{eq:defcorrelationlength},  
with an appropriate choice of $b$, yields a finite $\overline{\xi}$ which diverges at the transition. 
Furthermore, in the full scheme  the exponents $\tau,\beta,\nu_{\pm}$ are modified (their (non)-universality being hard to assess), 
e.g.\@ $3>\tau >2$, implying $\overline{S} \sim L^{3-\tau} \ll L$ at the transition. 
In particular, the rigorous Harris bounds \cite{chandran2015finite} on $\nu_\pm$ (violated by mean field) are satisfied, as it has to be.
As $\nu_-$ seems unrelated to avalanches, we expect $\nu_+\neq\nu_-$ also beyond mean field, but we have not been able to settle this numerically in the full scheme.

\textbf{Conclusion ---}
By studying the role of ergodic spots, we have argued that the critical MBL/ETH critical point is localized. 
The transition occurs as a quantum avalanche kicked off by the largest ergodic spots. 
We have identified the associated divergent length scales, all stemming from a divergent susceptibility to infinite ergodic spots. 
A simple mean-field flow equation allowed us to illustrate these features in an explicit and quantitative way.

\begin{acknowledgments} 
We are grateful to A. Scardicchio,  D. Huse and V. Khemani for helpful insights and discussions. 
FH benefited from the support of the projects EDNHS ANR-14-CE25-0011 and LSD ANR-15-CE40-0020-01 of the French National Research Agency (ANR).  
WDR acknowledges the support of the Flemish Research Fund FWO under grant G076216N. 
TT and WDR have been supported by the InterUniversity Attraction Pole phase VII/18 dynamics, geometry and statistical physics of the Belgian Science Policy. 
TT is a postdoctoral fellow of the Research Foundation, Flanders (FWO).
\end{acknowledgments}

\newpage

%
%
%
%
%
%
%
%
%
%
%
%
%

\newpage

\section*{Supplemental Material}

\subsection{The resonance condition}

It would be too strong to demand {the condition \eqref{eq:resonancecondition}} for {\em every} pair of states $(\eta ,\eta ')$. Some pairs of states are still allowed to be resonant in regions which are otherwise considered perturbatively coupled, as long as this type of resonances do not percolate. The latter is guaranteed if the probability for a given state $\eta$ to have any resonant partner $\eta'$ (that is, to violate the condition on $\mathcal G$) is much smaller than 1. 
On the other hand, it would be too weak to demand the condition only for typical {\em pairs} $(\eta ,\eta ')$. Indeed, if $d_s$, the number of states per site, is large, it can happen that almost every $\eta $ has a resonant partner $\eta '$, in which case the resonances will percolate,  while the number of resonant pairs $(\eta ,\eta ')$ is still a small fraction of all pairs. As an example, one can consider the interaction between many-body free-fermion eigenstates in a coarse grained picture, cf. \cite{ros2015integrals}. For a typical pair of such states, a quartic interaction vanishes, as the interaction only connects states that differ in at most 4 occupation numbers of single particle orbitals.

\subsection{The `rule of halted decay' in thermal regions}

This rule is invoked just before equation \eqref{eq: populistic} (main text). 
The setup is that we consider a coupling $V_I$ which has a fully diagonalized thermal spot in its domain and which extends at one end to a not yet diagonalized thermal spot as well, see the bottom picture in Fig.~\ref{Fig:halting}. 
Such couplings emerge only after a few steps (illustrated in the figure as well). In particular the thermal region should be non-perturbatively diagonalized and then one needs at least one perturbative elimination (on Fig.~\ref{Fig:halting}: elimination of $V_b$ acting on the coupling $V_a$) to produce $V_I$ ($V_I=[A_b,V_a]$). Our rule then states that the decay of the norm $\norm V_I \norm$ is halted inside the thermal (green) region.
\begin{figure}
 \begin{center}
{\includegraphics[width=9cm]{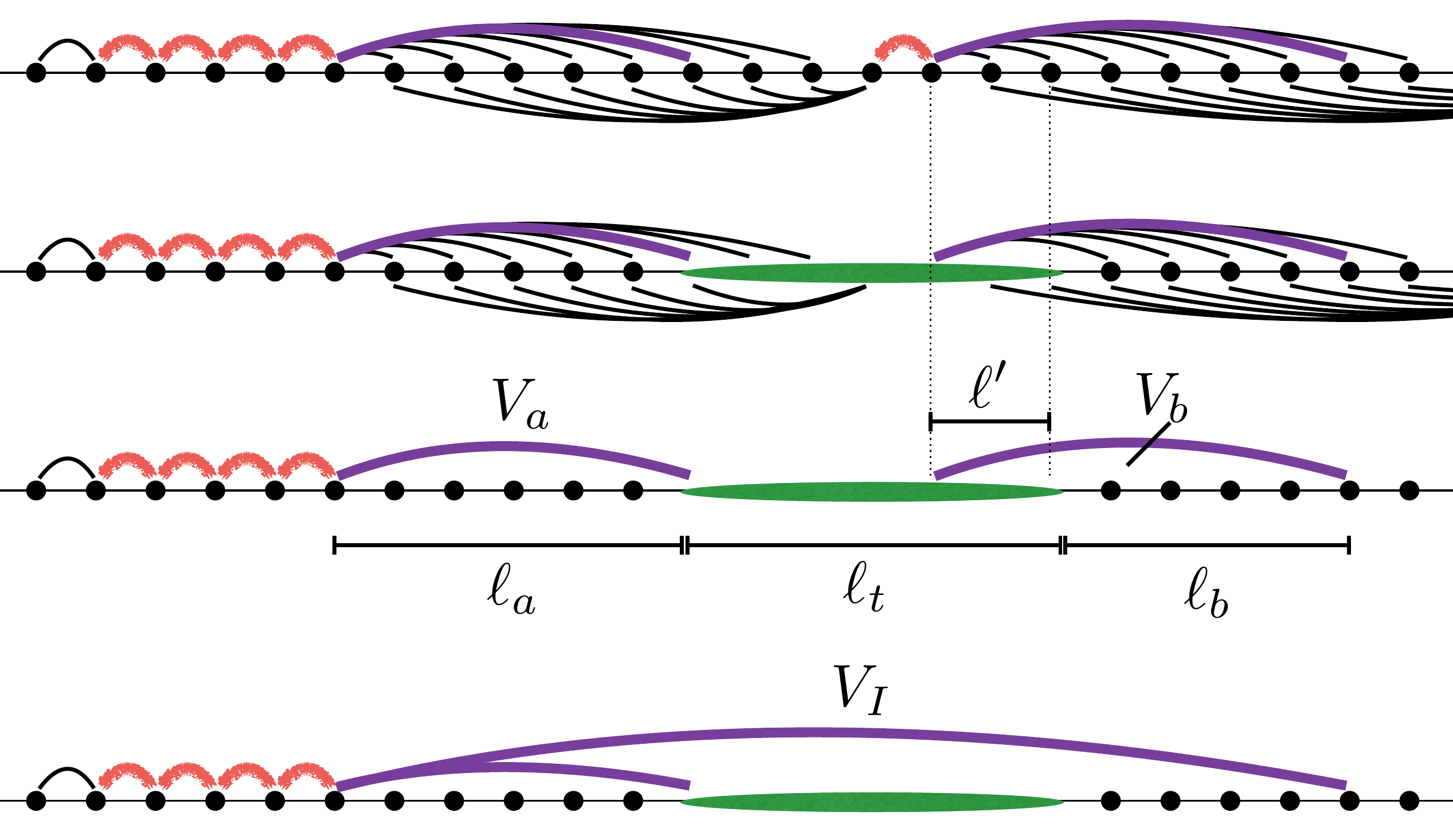}} 
\caption{The coupling $V_I$ as discussed in the text.  The color code is the same as that in Figure \ref{Fig:FirstBlob} (main text).  The green ellipse is the thermal spot embraced by the coupling. The three top figures illustrate how the coupling $V_I$ was constructed. {Note that throughout all steps, $V_a$ links to the active ergodic spot on the left, therefore it is not eliminated.}}
\label{Fig:halting}
\end{center}
\end{figure} 
Let us see that explicitly here, assuming that all couplings $V$ can be modeled by random matrix theory (RMT), {see below for some comments}.

 This in particular implies the following scaling between the norm $||V||$ of couplings and their typical matrix elements $m$: $||V|| \sim \sqrt{d} m $, with $d$ the dimension of the space on which $V$ acts, i.e.\ $d=\e^{s \ell}$ with $\ell$ the length of the region on which $V$ acts. We also assume that the norm of the couplings  $V_a$ and $V_b$ (see Fig.~\ref{Fig:halting}) decay with a localization length $\zeta$, i.e. $||V_a|| \sim e^{-\ell_a/\zeta}$ and $||V_b|| \sim e^{-(\ell'+\ell_b)/\zeta}$. It is useful to keep track of the operator norm since it is invariant under (non-perturbative) unitary conjugations. The length $\zeta$ characterizes the non-thermal regions on which the couplings acts. In the example of Fig.~\ref{Fig:halting}, we have $\zeta=\zeta_1$ because theses regions are completely free of resonances.  The size of the thermal spot, at the end of the non-perturbative rotation, is $\ell_t$, and $\ell'$ is the size of the region it has thermalized on its right.
The typical matrix elements of $V_I$ are evaluated at first order in perturbation theory in $V_b$ as (see details in \cite{thimotheelong})
\begin{equation}\label{eq: expression mI}
m_I \sim \frac{m_a m_b}{\delta_t} \sim \e^{s\ell_t} m_a m_b \, ,
\end{equation}
where $\delta_t \sim \e^{-s\ell_t} $ is the level spacing in the ergodic spot. In this calculation, we have neglected factors that are sub-exponential in the length of the bath (bandwidth...). Now we use RMT to get
$$m_b \sim \e^{-s(\ell_t+\ell_b)/2} e^{-(\ell'+\ell_b)/\zeta}, \qquad m_a \sim  \e^{-s(\ell_a+\ell_t)/2} e^{-\ell_a/\zeta}. $$ We find the length of the bath by invoking a  criticality condition, i.e.\ the bath expands until $\mathcal{G}\approx 1$, yielding
 $$ \e^{s\ell_t} \e^{-s\ell_t/2} e^{-\ell'/\zeta} \sim 1.$$ Hence we get $m_b \sim  \e^{-s\ell_t}  \e^{-s\ell_b/2} e^{-\ell_b/\zeta}$. We plug this into the expression \eqref{eq: expression mI} for $m_I$ and we use the RMT estimate $||V_I|| \sim  \e^{s(\ell_a+\ell_b+\ell_t)/2} m_I$ to obtain
\bea \label{SM:halting2}
|| V_I || \sim e^{-(\ell_a+\ell_b)/\zeta} \, ,
\eea
This corresponds to the rule stated in the main text because the total length $\ell_a+\ell_b+\ell_t$ is replaced by $\ell_a+\ell_b$ in the exponent. 

{As already indicated, our analysis depends on the use of RMT for the structure of perturbative couplings.  This might sound a bit surprising, and it has a priori nothing to do with resonances and ETH (unlike the use of RMT for \emph{resonant} couplings). 
We assume that the original (short-range) couplings are of random matrix type.  In \cite{thimotheelong} we showed that this random matrix assumption does not remain exactly valid for perturbative couplings: new couplings created during the RG acquire a structure stemming from the presence of energy denominators in their expression. However, this structure only manifests itself in the heavy tail of the distribution of matrix elements, while we can keep using RMT at each step to determine, e.g. the typical matrix elements of couplings in the next step. In contrast, the norm of such operators is now dominated by outliers, and thus the deviation from RMT shows up in the relation between norm and typical matrix elements. 
The main consequence for our RG is that \eqref{SM:halting2} is modified into $|| V_I || \sim e^{-(\ell_a+\ell_b)/\zeta -\ell_t/\zeta_t} $, with $\zeta_t = - 2 \zeta_c$. }
Hence in principle the 'rule of halted decay' is modified, it even becomes a 'rule of increase' and the mean-field flow equation \eqref{eq: populistic} should be replaced by $\loct_{k+1}^{-1}= (1-\rho_k)\loct_{k}^{-1}+\rho_k \zeta_t^{-1}$. However, it can easily be checked that our conclusions about the critical behavior are independent of $\zeta_t$ as long as {$1/\zeta_t<1/\zeta_c$} (ensuring that treated thermal regions are supercritical and can lead to instability). In the letter we therefore simplify the analysis by making the intuitive choice  $1/\zeta_t=0$.

\subsection{The mean field model}

\subsubsection{Justification} 
 \setcounter{paragraph}{0}
 {
Mathematically speaking, our mean-field model is defined by the expressions 
$$\loct_1(\epsilon)= - 1/\log(\epsilon/K),\qquad \rho_k\equiv\epsilon^k (k+2\ell_k/a) $$
 together with the flow equation \eqref{eq: populistic}, i.e.\
 \begin{equation} \label{eq: populistic repeated}
\loct_{k+1}^{-1}= (1-\rho_k)\loct_{k}^{-1},
 \end{equation}
and the relation \eqref{eq: def lc}, i.e.\ 
\begin{equation} \label{eq: def lc repeated}
\ell_k = k \tfrac{s a}{2}
\left(\tfrac{1}{\loct_k}-s\right)^{-1}.
\end{equation}
The expression for $\loct_1(\epsilon)$ is natural but not fundamental. In practice, we view it as a way to trace the fate of the parameter $K$, allowing us to witness non-universality. 
The equation \eqref{eq: def lc repeated} is well-justified by numerics in \cite{luitz2017} provided that $\zeta_k$ is indeed the effective localization length in a region of length $\ell_k$ around the bare spot. }

{Roughly speaking, there are two major assumptions underlying our mean-field analysis. 
 The first assumption consists in neglecting situations where a bare spot invades another (larger) bare spot while it is melting.  Indeed, in the flow equation \eqref{eq: populistic repeated},  we are pretending that the environment of a bare spot of order $k$ consists of fully treated (i.e.\ diagonalized) spots of smaller order.  
 The second assumption is very much related: 
In a stretch of length $\ell$  around the bare spot of order $k+1$, we can define the volume covered by thermal spots of order $k$. The average of this volume is given by $\rho_k \ell$.  Yet, if $\ell$ is of order of $\ell_{k+1}$, this volume is zero with large probability, i.e.\ it is very rare to find a spot of order $k$ there. However, in equation \eqref{eq: populistic repeated}, we neglect the fluctuations of this volume on the length scale $\ell=\ell_{k+1}$ and replace it by its average.}

The analysis and numerics in \cite{thimotheelong} confirm that the full scheme, where these approximations are not made, differs only mildly from our mean-field analysis, without altering the qualitative picture. An obvious expected difference is that the critical exponents are modified. However, the most spectacular difference is that, in the full scheme, the subcritical distribution of thermal spot sizes $p(\ell)$ is decaying as a stretched exponential rather than an exponential. This means that the averaged localization and correlation lengths ($\loca, \xia$) in fact diverge in the entire localized region.  However, as also indicated in the Letter, it is more natural to redefine these notions so as to get finite lengths (except at the transition). This can be done either by defining them as the scale at which the critical size distribution of spots is cut-off below criticality (regardless of the nature of this cutoff; exponential $e^{-\ell/\loca}$ or stretched exponential$e^{-(\ell/\loca)^b}$, with $0<b<1$), or, by replacing the inverse length $1/|I|$ in \eqref{def: loc lengths} by  $1/|I|^b$.

\subsubsection{Analysis of the flow equations} 
\setcounter{paragraph}{0}
Here we analyze the flow equations for the running localization length $\zeta_k$ and `collar' length $\ell_k$ as defined by Eq.~\eqref{eq: populistic repeated}-\eqref{eq: def lc repeated}. Introducing $\alpha_k = 1/\zeta_k$ these are rewritten as
\bea \label{eq:flow1}
&& \alpha_{k+1} = (1-\rho_k) \alpha_k \\
&& \ell_k = k(\tfrac{s a}{2}) (\alpha_k- \alpha_c)^{-1} \, , \label{eq:flow2}
\eea
with $\alpha_c=s$ and $\rho_k = \epsilon^k (k + 2 \ell_k /a)$.

\paragraph{Scaling of collars in the MBL phase} In the MBL phase $\epsilon<\epsilon_c$, the inverse (typical) localization length converges as $\lim_{k \to \infty} \alpha_k  = \alpha_\infty > \alpha_c$. The collar length grows linearly with the size of the bare spot as
\bea \label{eq:LinearCollar}
\ell_k \simeq \kappa_1 k \quad , \quad \kappa_1 = (\tfrac{s a}{2}) (\alpha_\infty - \alpha_c)^{-1} \, .
\eea

\paragraph{Scaling of collars at the critical point}
At the critical point $\epsilon=\epsilon_c$, $\lim_{k \to \infty} \alpha_k = \alpha_c$. Writing $\alpha_k = \alpha_c + \delta \alpha_k$ with $ \delta \alpha_k>0$ and $\lim_{k \to \infty} \delta \alpha_k = 0$, the collar length now grows superlinearly as 
$$
\ell_k \simeq (\tfrac{s a}{2}) \frac{k}{\delta \alpha_k } \, .
$$
Inserting in \eqref{eq:flow1} and expanding in $\delta \alpha_k$ leads to
\bea \label{Eq:deltaalphadisc}
\delta \alpha_k (\delta \alpha_{k+1} - \delta \alpha_k) \sim - s^2 k \epsilon_c^k .
\eea
Hence the inverse localization length converges as 
\bea
\delta \alpha_k \sim \sqrt{\frac{-2}{\log(\epsilon_c)}} s \sqrt{k \epsilon_c^{k}} \, ,
\eea
and collars grow exponentially as
\bea \label{eq:ExponentialCollar}
\ell_k = \frac{a}{2\sqrt{-2 \log(\epsilon_c)}} \sqrt{k \epsilon_c^{-k}} \, .
\eea

\paragraph{Insight from a continuum approximation} To get more insight into the behavior of the system close to the transition it is useful to consider a continuum approximation of \eqref{Eq:deltaalphadisc} taken for $\epsilon \simeq \epsilon_c$ as 

\bea \label{Eq:deltaalphacont}
\partial_k \delta \alpha_k^2 =-2 s^2k \epsilon^k \, .
\eea

Assuming first that we are in the MBL phase $\epsilon < \epsilon_c$ and integrating \eqref{Eq:deltaalphacont} from $k=1$ to $k=+\infty$ leads to 
\be \label{Eq:Calculdelta}
(\alpha_{\infty}(\epsilon)-\alpha_c)^2 = (\alpha_1(\epsilon)-\alpha_c)^2  -2 s^2   \int_{k=1}^\infty k \epsilon^k  dk \, .
\ee
Expanding around the critical point and using that the right hand side of \eqref{Eq:Calculdelta} is a regular function of $\epsilon$ that is $0$ at the critical point, we obtain that $(\alpha_\infty(\epsilon) - \alpha_c) \sim_{\epsilon \sim \epsilon_c^-} \sqrt{\epsilon_c- \epsilon}$. Thus we find {
\bea \label{eq: scaling zeta}
\zeta_c - \zeta \sim (\epsilon_c -\epsilon)^{\nu_-}
\eea
with an exponent $\nu_- = 1/2$.} Above the critical point we can estimate the scaling of $k_{*}(\epsilon)$, the size of the first resonances that cause delocalization, now using that $\delta \alpha(k_{*}(\epsilon))=0$. We get
\bea
\int_{k=1}^{k_{*}(\epsilon)} k \epsilon^k dk = \frac{1}{2s^2} (\alpha_1(\epsilon)-\alpha_c)^2  \nonumber \, .
\eea
 Hence, using $k_{*}(\epsilon_c)=+\infty$,
\bea 
&& \int_{k=1}^{k_{*}(\epsilon)} k \epsilon^k dk - \int_{k=1}^{+\infty} k \epsilon_c^k dk  \nn
&& = \frac{1}{2 s^2}  \left( (\alpha_1(\epsilon)-\alpha_c)^2  - (\alpha_1(\epsilon_c)-\alpha_c)^2  \right) \nonumber \, .
\eea
Expanding around $\epsilon_c$ we get $\int_{k_{*}(\epsilon)}^{+\infty}  k \epsilon_c^k dk - \int_{1}^{+\infty} k^2 \epsilon_c^{k-1}(\epsilon-\epsilon_c) dk \sim \frac{1}{s^2} \alpha_1'(\epsilon_c)(\epsilon-\epsilon_c)(\alpha_1(\epsilon_c)-\alpha_c)  $. Hence $k_*(\epsilon) \epsilon_c^{k_*(\epsilon)}$ is of order $\epsilon-\epsilon_c$ and we get {
\bea \label{Eq:Divergencekmin}
k_{*}(\epsilon)  \sim \frac{\nu_+}{\log(\epsilon_c)} \log(\epsilon-\epsilon_c) \, ,
\eea
with $\nu_+ = 1$.}

\paragraph{Finite size scaling exponents} The exponents $\nu_{\pm}$ introduced in the Letter to characterize the divergence of the lengthscales $L_{\pm}$ controling the finite size scaling of $p(\epsilon,L)$ are thus evaluated in the continuum approximation as $\nu_-=1/2$ and $\nu_+ = 1$. Analyzing numerically the recursion equations \eqref{eq:flow1}-\eqref{eq:flow2} we find that the qualitative behavior \eqref{eq: scaling zeta}-\eqref{Eq:Divergencekmin} is correct but that the value of $\nu_+$ and $\nu_-$ differ from those predicted by this calculation. The use of the continuum approximation \eqref{Eq:deltaalphacont} thus appears incorrect (that can easily be checked) and the discreteness of the flow equations leads to corrections to these scaling exponents. We find numerically that both $\nu_+ $ and $\nu_-$ appear to be non-universal: they depend explicitly on the choice of the function $\zeta_1(\epsilon)$ (bare localization length in terms of the resonance probability). Taking $\zeta_1(\epsilon) = - 1/\log(\epsilon/K)$ we find that as $K$ grows $\epsilon_c$ grows and $\nu_-$ and $\nu_+$ approach the values $1/2$ and $1$ predicted by the calculation, while in general we find $1/2<\nu_-<1$ and $1<\nu_+ <2$. We note that these mean-field predictions for $\nu_\pm$explicitly break the rigorous Harris-bound of \cite{chandran2015finite}. {That only appears as a caveat of our mean-field analysis: numerics on the full scheme \cite{thimotheelong} leads to exponents in agreement with  \cite{chandran2015finite}.}

\paragraph{Critical regime} The transition point is MBL but characterized by the anomalous (exponential) growth of the collar lengths \eqref{eq:ExponentialCollar}. In a finite system of size $L$, slightly below the transition point $\alpha_{\infty}>\alpha_c$, one can also observe this anomalous growth if $L \leq L_-^{{\rm crit}} \sim \epsilon^{k_-}$ where $k_-$ is obtained by matching the scaling of collars in the MBL regime \eqref{eq:LinearCollar} and at the critical point \eqref{eq:ExponentialCollar}, i.e.
$$
k_- \frac{sa}{2}(\alpha_{\infty} - \alpha_c)^{-1} \sim \frac{a}{2\sqrt{-2 \log(\epsilon_c)}} \sqrt{k_- \epsilon_c^{-k_-}} \, ,
$$
we get
\bea
k_- \sim \frac{2 \log(\alpha_{\infty}-\alpha_c)}{\log(\epsilon_c)} \, ,
\eea
and $L_-^{{\rm crit}}$ diverges algebraically as $L_-^{{\rm crit}}\sim \frac{1}{(\alpha_\infty- \alpha_c)^2} \sim (\epsilon_c-\epsilon)^{-2 \delta}$. {This also implies that below the critical point the thermal spots distribution $p(\ell)$ display the same power law as at criticality $p(\ell) \sim \ell^{-3}$ up to a scale $\ell \sim \epsilon^{-k_-/2} \sim \loca$.} Above the transition point and in a finite system of size $L\leq L_+^{{\rm crit}}  = L_+ = \epsilon^{—k_*(\epsilon)} \sim (\epsilon-\epsilon_c)^{- \nu_+}$ with $\nu_+ = -\tilde{\nu}_+ \log(\epsilon_c)$, one also expects that MBL systems exhibit this anomalous growth of collars. In a finite system this extends  {the localized critical regime on {\it both} sides of the transition}.

\subsection{Calculation of correlation lengths} 

\paragraph{General theory}
Consider the vicinity of a full thermal region (bare spot + melted region) after it has been diagonalized, see Fig.~\ref{Fig:spot}.
The Hamiltonian is given by  
$$
H = H_{\rl}+H_{\rr}+ H_{\rE}+ V_{\mathrm{E-l}}+V_{\mathrm{E-r}}.
$$
Here, $H_\rE$ describes the thermal region with $\ell_\rE$ spins, whereas $H_\rl,H_\rr$ act on the neighboring spin on the left/right, which we assume to never have been involved in a resonance.    
The couplings $V_{\mathrm{E-l}},V_{\mathrm{E-r}}$ are perturbative (otherwise the l/r spins would be melted as well), but barely so, as they are the first spins not to melt. Hence $\mathcal G_{\mathrm{E-l}},\mathcal G_{\mathrm{E-r}} \sim 1$. Let us investigate the implication of this for a typical eigenstate on $\rl+\rE+\rr$, which we decompose as
 $
 \Psi = \sum_{s_\rl,s_\rr,b}   c(s_\rl,s_\rr,b) \str s_\rl \rangle \str s_\rr \rangle \str b\rangle,
 $
 with $\ket{s_\rl}$, $\ket{s_\rr}$ and $\ket{b}$ the eigenstates of $H_\rl  $, $H_\rr$ and $H_\rE$. Let $N$ be the number of configurations $(s_\rl,s_\rr,b)$ on which the coefficients $c(s_\rl,s_\rr,b)$ are mainly supported and let $\rho=\mathrm{tr}_\rE |\Psi\rangle \langle \Psi|$ be the reduced density matrix of $\rl+\rr$. Simple considerations {(see below) lead to the following properties for $N,\, \rho$ and $\mathrm{Cor}(O_\rl,O_\rr)$ as a function of ${\cal G} =\mathcal G_{\mathrm{E-l/r}}$}, {(where the middle row applies to our case)}:\\
\begin{center}
  \begin{tabular}{| c | c |c |c| }
    \hline
   $ \mathcal G$ & $N$  & $\rho$ & $\mathrm{Cor}(O_\rl,O_\rr)$  \\ \hline \hline
    $ \ll 1 $ & $1$  & product(pure) & $\approx 0  $    \\ \hline
        $ \approx 1 $ & $O(1)$  & no structure & $ O(1)  $    \\ \hline
                $ \gg 1 $ & $O(e^{s l_\rE})$  & product(mixed) & $ \approx 0  $    \\ \hline 
  \end{tabular} \, .
\end{center} 
This highlights the special role of the critical case $\mathcal{G} \approx 1$, relevant to the subsystem $\rl+\rE+\rr$. Hence we find $\mathrm{Cor}(O_\rl,O_\rr)\approx 1$.
\begin{figure}
\begin{center}
{\includegraphics[width=9cm, height=3.5cm]{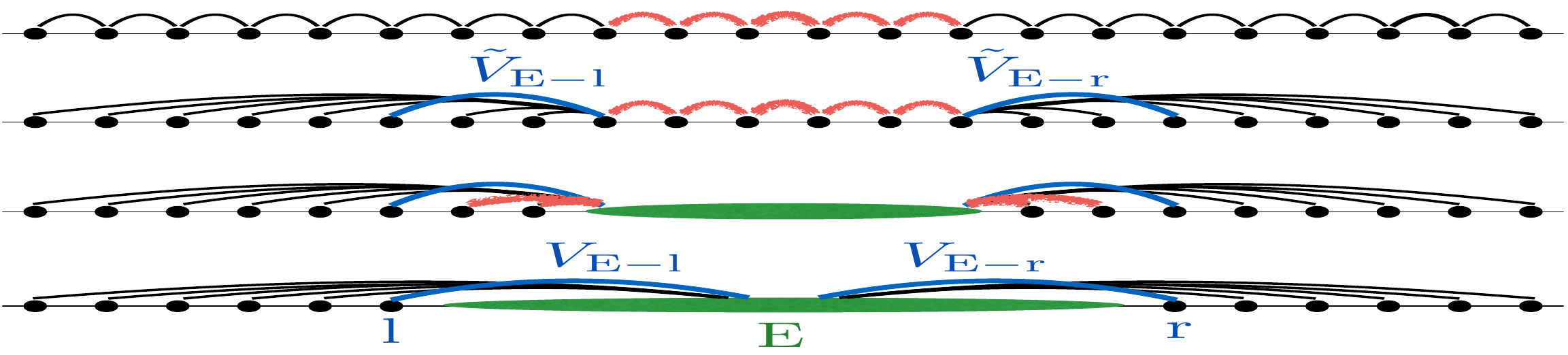}} 
\caption{
Growth of an ergodic spot, according to our scheme, see Fig.~\ref{Fig:FirstBlob} (main text) for the color code. The blue couplings will eventually link the spot to its neighbors.
}\label{Fig:spot}
\end{center}
\end{figure} 
The correlator is $O(1)$ with the probability of a thermal region spanning the considered interval, as in the calculation of $\loca$ above. Hence we find
$
\xia \sim \loca
$. 

Let us now substantiate the claims in the above table. 
In the absence of coupling (${\cal G}_{\mathrm{E-l}} = {\cal G}_{\mathrm{E-r}}=0$), any eigenvector $\Psi$ is a product state and $c$ is non-zero for a single combination $(s_\rl',s_\rr',b')$. In the case where $s_\rl,s_\rr,b$ are resonant ($\mathcal G_{\mathrm{E-l}},\mathcal G_{\mathrm{E-r}} > 1$ ) we get $\Psi$ be applying a random unitary to $\str s_\rl \rangle \str s_\rr \rangle \str b\rangle$ and then there is a number $N =2^{\ell_\rE} \gg 1$ of non-zero $c$'s.  
 In the critical case that is considered here $c(s_\rl,s_\rr,b)$ is non-zero for a `small' number of $(s_\rl,s_\rr,b)$: $N=O(1)$. Indeed, since we are  in the perturbative case (though barely so),
$$\Psi \simeq e^{i (A_{\mathrm{E-l}}+A_{\mathrm{E-r}})} \str s_\rl' \rangle \str s_\rr' \rangle \str b'\rangle, \qquad \text{for some $(s_\rl',s_\rr',b')$},
$$
and the coefficients $c(s_\rl,s_\rr,b)$ can be evaluated by expanding in powers of $A_{\mathrm{E-l}}, A_{\mathrm{E-r}}$. Correlations  between the spins $\rm l,\rm r$ appear in second order (first order in both $A_{\mathrm{E-l}}$ and  $A_{\mathrm{E-r}}$). The corresponding contribution to $c(s_\rl,s_\rr,b) $ is 
$$\sum_{b''} \tfrac{\braket{ s_\rl b |V_{\mathrm{E-l}}| s_\rl' b'' }}{E(s_\rl)+E(b)-E(s_\rl')-E(b'')} \tfrac{\braket{ s_\rr b'' |V_{\mathrm{E-r}}| s_\rr' b' }}{E(s_\rr)+E(b'')-E(s_\rr')-E(b')}.$$
Having fixed $(s_\rr,s_\rl,s_\rr',s_\rl',s_b)$ and since $V_{\rE-\rr}$ is critical, the second factor in the sum is of order ${\cal G}_{\mathrm{E-r}} \sim 1$ for a few $b''$ that minimize the denominator and much smaller otherwise. Since $V_{\rE-\rl}$ is also critical, the overall sum is thus of order  ${\cal G}_{\mathrm{E-l}}  \times {\cal G}_{\mathrm{E-r}} \sim 1$ for a few $b$. On the other hand for typical $b$, $b''$ cannot be chosen in such a way that both denominators are small, and in that case the sum is exponentially small in the length of the bath (since the denominators are typically $O(1)$ while the numerators are typicall exponentially small in the length of the bath). Eigenstate correlations of the type \eqref{def: correlation} can be read off from the reduced density matrix
 $$
\rho(s_\rl,s_\rr,s'_\rl,s'_\rr)=  \sum_b c(s_\rl,s_\rr,b) \bar{c}(s'_\rl,s'_\rr,b) .
$$
In the absence of coupling $\rho$ is diagonal, but that is also the case in the strong coupling case $N \simeq 2^{\ell_E} \gg 1$:  from the law of large numbers and the central limit theorem we obtain that $\rho=1/4+\mathcal O(2^{-\ell_E/2})$.  This is of course just a contrived way to express the typicality in random matrix theory. In the critical case $N=\mathcal{O}(1)$, these consideqrations do not apply and the density matrix $\rho$ is a random $4\times 4$ operator without any specific structure. This implies that the correlation between the spins is of order $1$.

\paragraph{End-to-end correlations at $\epsilon>\epsilon_c$}

{We estimate, through a particular mechanism, the probability that the correlator between endpoints in a supercritical systems of length $L$ is $O(1)$. First, the system needs to be in the localized phase, which costs $p_+(L)\sim \e^{-L/L_+}$, the probability that there is no bare spot of order $k_*$. Furthermore, conditioned on this, we need enough resonances so that there is a thermal spot covering almost the  whole system. Using the properties of spots at criticality, we learn that the length thermalized by a bare spot of order $k_*$ (actually $k_*-1$, but the difference is irrelevant)  is $\epsilon_c^{k_*/2}\sim \sqrt{L_+}$. The required number of such spots is hence 
$$
N \sim L/{\sqrt{qL_+}},
$$
with $q$ some non-divergent length scale. This leads to the probability for near-thermalization
$$
p \sim   \epsilon_c^{k_* N}   \e^{-\epsilon_c N}
$$
In this expression, the second factor is due to the conditioning on the absence of spots of higher order. Clearly, it is negligible compared to the first factor.  We get hence, neglecting constants,
$$
p \sim  (1/L_+)^{\tfrac{L}{\sqrt{L_+}}} \sim \e^{-\log L_+ \frac{L}{\sqrt{L_+}}} 
$$
This is smaller than $p_+(L)$ and hence we obtained here that the end-to-end correlation length scales as $(\log L_+)^{-1} \sqrt{L_+}$.
This is merely a  lower bound since we did not investigate alternative mechanisms for slow decay.

}

%
%

\subsection{The scheme in physical time}
{
Though our theory is formulated primarily as a scheme for determining the eigenstates, we can also view it as encoding the dynamics. 
To do so, we discard local phase oscillations and we focus on the dissipative dynamics. Then, any perturbative step (elimination of couplings) does nothing dynamically, while the fusion of a region means that this region thermalizes. The rate of thermalization is determined from the resonant coupling $V_I$ that was associated to this region, by the 
following Fermi Golden Rule formula
$$
\Gamma =  \frac{m^2_I}{\delta} 
$$
where $\delta$ is the level spacing in the stretch $I$ and, as before, $m_I$ is a typical matrix element of $V_I$. If the fused region is connected by more than one resonant couplings, then we should take the minimal rate. 
Note that one could make the scheme even more theoretical and detailed by insisting that we are building up the structure factors that enter in the ETH prescription. In that view, we are hybridising states that coincide outside the fused region and that lie within an energy window $\Gamma$ of each other (see also \cite{de2017stability} for much more details on such a procedure). 
}
\paragraph{Local thermalization rates}
{
Back to the scheme in physical time.  It is clear that in general, the scheme will proceed towards fusing larger and larger regions, which corresponds to ever longer timescales. Consider a spin  thermalized by a bare spot at distance $\ell$. Then $||V_I|| \sim \e^{-\ell/\loct}$, and by using the random matrix relation $||V_I||=\sqrt{d_I}m_I$ and $\delta d\sim 1$, we find 
$$
 \Gamma_\ell \sim  \Gamma_0 \e^{-2\ell/\loct}
 $$
 with $\Gamma_0$ some reference rate. We see hence that close to criticality, with many eventually thermal sites at various distances $\ell$,  there is a wide distribution of thermalization rates throughout the chain.  As indicated in the main text, the typical thermalization time $t_+$ is given by
 $$
t_+ \sim   \Gamma^{-1}_{\ell=L_+}. 
 $$
It diverges rapidly, as  $\epsilon\to \epsilon_c$, and so does also the width of the distribution, being itself proportional to $t_+$. }

\paragraph{Transport}
{
Let us now discuss how to identify transport within this picture.  In principle, to get transport across a chain, one needs every link in the chain to be affected by dissipative processes. From this, one might express the resistance across a chain as a sum of resistances, each of which is roughly given by the local inverse rate $\Gamma^{-1}$.  
On scales smaller than $L_+$, the picture is actually much simpler as the sum will be dominated by the smallest $\Gamma$. Hence on scales $\ell \ll L_+$, one finds a completely anomalous transport law, with resistance increasing exponentially with length. On scales $\ell \gg L_+ $, one sees multiple spots of order $k_*$, spaced approximately at intervals of length $L_+$, and it becomes important to take the sum of $\Gamma^{-1}$.  If all such spots were placed exactly at intervals of length $L_+$, the resistance would grow linearly with length, thus yielding a normal (diffusive) transport law, albeit with a diverging resistivity as $\epsilon\to\epsilon_c$.   Of course, in reality the explosive spots are not uniformly spaced and anomalously long intervals between them play the role of blocking regions, exactly as it is in the standard argument for subdiffusive transport with strong disorder (or subballistic entanglement spreading), see \cite{lev2015absence,luitz2016extended,gopalakrishnan2016griffiths,luitz2017ergodic,vznidarivc2016diffusive,nahum2017dynamics}.  
Let us spell this out more explicitly.  The probability that a certain stretch of the chain of large length $\ell$ is free of explosive spots is given by $\e^{-\ell/L_+}$. The resistance of such a stretch is $\sim \e^{2\ell/\loct}$. If $1/L_+ >2/\loct$, i.e.\ far enough from the transition, then the resistance is not dominated by rare regions and the transport is diffusive. In the opposite case $2L_+ \geq \loct$, the resistance $R$ is dominated by the largest stretch free of explosive spots and it is of order
$$
R(L) \sim L^{\frac{2L_+}{\loct}}  \sim L^{C(\epsilon-\epsilon_c)^{-\nu_+}}
$$
Hence, the 'power' governing the subdiffusion diverges upon approaching the transition.
}

\subsection{Comparison with previous RG schemes}
 
{Several RG schemes for the MBL transition have been studied before \cite{vosk2015theory,potter2015universal,dumitrescu2017scaling}. The main feature that distinguishes our approach is our focus on the 'avalanche' instability, understood as the ability of a single spot to thermalize a full chain. This allows us to develop a theory that reduces the reliance on numerics.  For example, in \cite{potter2015universal}, it is remarked explicitly that avalanches are irrelevant for the critical behaviour.  In \cite{vosk2015theory}, the authors explicitly restrict the RG rules so as to avoid avalanches, which are deemed unphysical, see page 4 of \cite{vosk2015theory}.

%
%
%
%
{
The RG scheme of \cite{dumitrescu2017scaling} is close in spirit to our work as it also uses random matrix estimates to deal with resonant spots. From a theoretical point of view, the main difference with our scheme is the following: In our scheme the localization length (LL) at large scales is renormalized and it is strictly larger than the 'bare' LL, defined as the LL of a resonance-free chunk of material. This renormalization emerges through perturbative rotations, leading to the 'rule of halted decay' (see above) and to the flow of the LL, for example captured by the equation \eqref{eq: populistic repeated}.
 This is in sharp contrast with  \cite{dumitrescu2017scaling} where couplings between all spins are initially generated with the LL $x_0$. The couplings are updated later on, but only when resonant clusters are fused (in which case the update rule is identical to ours).  
 To be very concrete, consider the coupling $V_I$ in the lower picture of Figure \ref{Fig:halting}, coupling the red spot on the left with the far right, and embracing a treated thermal region (depicted in green). 
 %
%
%
 In this case, a typical matrix element of $V_I$ according to \cite{dumitrescu2017scaling} is given by $\exp(-\ell/x_0)$, where $\ell=\ell_a+\ell_b+\ell_t$ is the full length of the stretch $I$. Most importantly, this expression is independent of the length of the green thermal region.  According to our 'rule of halted decay', a typical matrix element of $V_I$ would be given by
$ \e^{-s\ell_t/2} \exp(-[\ell-\ell_t]/x_0) $. 
 This amounts to a larger effective LL because necessarily  $x_0 \leq \tfrac{1}{s} \leq \tfrac{2}{s} $, cf.\ the discussion in the main text.
 To check this calculation, one should keep in mind that $x_0$ and $\zeta$ should not be confused, since they are LL for matrix elements and norms, respectively.

{This renormalization of the LL seems important, in particular at criticality. Indeed, the origin of the critical behavior is that the flow equations become singular when the renormalized LL approaches its critical value at large scales.  It is exactly the fact that the large scale LL hits the critical value $1/s$, while the small scale LL is still subcritical, which enables us to have a localized critical point. }

{Finally, the results of our numerical analysis are also quite different from  \cite{dumitrescu2017scaling}. Indeed, in  \cite{dumitrescu2017scaling} the critical point can be thermal (although with a small probability).   This also shows up in the distribution of entanglement and cluster sizes. For example, it is reported in  \cite{dumitrescu2017scaling} that the typical cluster length also diverges at the transition from the MBL side. Such a behavior is impossible in our scheme. }

\bibliographystyle{rsc}
\bibliography{loclibrary}

\begin{mcitethebibliography}{58}
\bibitem[ord()]{ordernote}
It is easy to see that one gets nonsense by eliminating couplings in the wrong
  order. For example, assume that we would always eliminate first the couplings
  that touch a resonant spot. Then such couplings would almost always be
  resonant and a single resonant spot would almost always thermalize the full
  chain.\relax
\mciteBstWouldAddEndPunctfalse
\mciteSetBstMidEndSepPunct{\mcitedefaultmidpunct}
{}{\mcitedefaultseppunct}\relax
\EndOfBibitem
\bibitem[not()]{notenorm}
The operator norm is defined as $\norm O \norm=\sup_{\psi: \norm \psi \norm_2
  =1 } \norm O\psi\norm_2$ where $\psi$ runs over vectors in Hilbert space and
  $\norm \cdot \norm_2$ is the usual {$L_2$} norm.\relax
\mciteBstWouldAddEndPunctfalse
\mciteSetBstMidEndSepPunct{\mcitedefaultmidpunct}
{}{\mcitedefaultseppunct}\relax
\EndOfBibitem
\end{mcitethebibliography}

\end{document}